\documentclass[twocolumn,showpacs,preprintnumbers,amsmath,amssymb]{revtex4}
\usepackage{graphicx}% Include figure files
\usepackage{dcolumn}% Align table columns on decimal point
\usepackage{bm}% bold math
\usepackage[colorlinks]{hyperref}
\begin{document}

%\preprint{APS/123-QED}

\title{Epidemic spreading on undirected and directed 
scale-free networks with correlations}% Force line breaks with \\

\author{Yukio Hayashi}
%\author{}
\affiliation{%
Japan Advanced Institute of Science and Technology,\\
Ishikawa, 923-1292, Japan
}%

%\affiliation{

\date{\today}% It is always \today, today,

\begin{abstract}
Many complex networks have a common structure 
with power-law degree distributions, 
however the details such as the degree-degree 
correlations are different in social, technological, and biological 
systems.
We numerically investigate the epidemic spreading on the 
network with a variety of correlations: 
the assortative, uncorrelated, and disassortative mixings.
In a simulation for the mean-field-like susceptible-infected-recovered 
model, we observe different epidemic behavior according to the types of 
correlations particularly in a directed network.
Our results suggest that the 
assortative connections between nodes with similar degrees enhance the
 epidemic spreading more significantly than the uncorrelated and 
disassortative connections between cooperative nodes with high and low 
degrees.
\end{abstract}

\maketitle

%\section{\label{sec:level1}First-level heading:\protect\\ The line
%break was forced \lowercase{via} \textbackslash\textbackslash}
%\subsection{\label{sec:level2}Second-level heading: Formatting}

Self-organized complex networks have attracted a great attention to 
statistical physicists, computer scientists, and mathematical
biologists, 
since many empirical studies have revealed the fact that 
a structure is commonly found in 
social, technological, and biological networks.
The structure is called scale-free (SF) \cite{Barabasi99} 
and follows a power-law distribution 
$P(k) \sim k^{-\gamma}$, $2 < \gamma < 3$, for the number of nodes with
degree $k$.
Since 
the heterogeneous characteristics of the SF network is crucial for 
the robustness of connectivity against failures \cite{Albert00a}, 
the efficiency of information delivery \cite{Cancho01}, 
and the spread of epidemic disease transmitted by means of social or
sexual contacts, e-mails, Internet, and so forth \cite{Satorras04},
we expect that the evolutional mechanisms and the structural
properties of the SF network \cite{Albert00a} are useful for improving 
the efficiency and the robustness 
of power supply, communication, and economy systems.

On the other hand, 
recent studies classify networks according to quantities of connectivity 
correlations of nodes with their neighbors \cite{Vazquez02};
social networks tend to have assortative connections between peers with
similar degrees \cite{Capocci03}\cite{Newman02c}, 
while technological or biological networks tend to have disassortative ones 
between those nodes with high degrees, namely hubs, and those with
low degrees \cite{Vazquez02}\cite{Newman03}.
The properties 
for the epidemic incidence \cite{Moreno03} and for the 
percolation \cite{Vazquez03a} have been compared 
in the considered forms of correlations 
defined as a weighted combination of the uncorrelated (or arbitrary 
correlated) term and the fully assortative term: 
a fully assortative connection allows only two nodes with the same
degree to be connected.
However the relation of the results stated above 
to the evolutional mechanisms of network is still unclear.
Only a few analytical forms of correlations
have been derived from 
the tree model \cite{Krapivsky01} and the configuration 
model restricted such that at most one link exists 
between any pair of nodes \cite{Park03}.
Although the configuration model produces disassortative mixing,
a non-trivial distribution of the desired degrees 
must be given for wiring in advance.
Also, difficulties in estimating the conditional probability of 
degree-degree connectivity 
even from empirical data has been 
pointed out \cite{Dorogovtsev03}.
Apart from the evolutional mechanisms, 
a numerical simulation has shown that 
the introduction of assortative hub-hub connections between different
local areas on a lattice has the effects to shorten 
the average delivery time and to enlarge the spread of
infection \cite{Singh03}.
However, neither theoretical solutions 
nor numerical studies for the epidemic
behavior have been reported except for the above special forms of 
correlations.

We study epidemic spreading on the SF network with a variety of 
assortative, disassortative, and uncorrelated connections,
which are not specialized in social, technological, or biological systems.
First, we review two growing network models: 
one is called the duplication-divergence model 
\cite{Vazquez03b}, which is equipped with a control parameter of
connectivity correlations 
between the assortative and the disassortative mixings, 
and the other is called the directed growing model, 
in which the existence of 
correlations is suggested \cite{Dorogovtsev03}.
Then, we will estimate the conditional probability of 
degree-degree correlations from the average realizations of the growing
network models. 
This estimation method can be applied to other network models.
By using the estimated probability,
we numerically investigate the epidemic behavior 
for the mean-field-like equations of the 
susceptible-infected-recovered/removed (SIR) model.
We find 
different epidemic behavior 
according to the types of correlations.

\begin{figure}[htb]
  \begin{minipage}[htb]{.47\textwidth}
    \includegraphics[height=25mm]{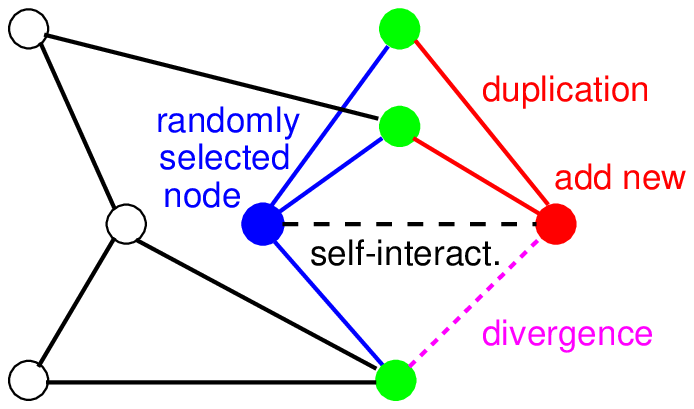}
    \begin{center} (a) \end{center}
  \end{minipage} 
  \hfill 
  \begin{minipage}[htb]{.47\textwidth}
    \includegraphics[height=30mm]{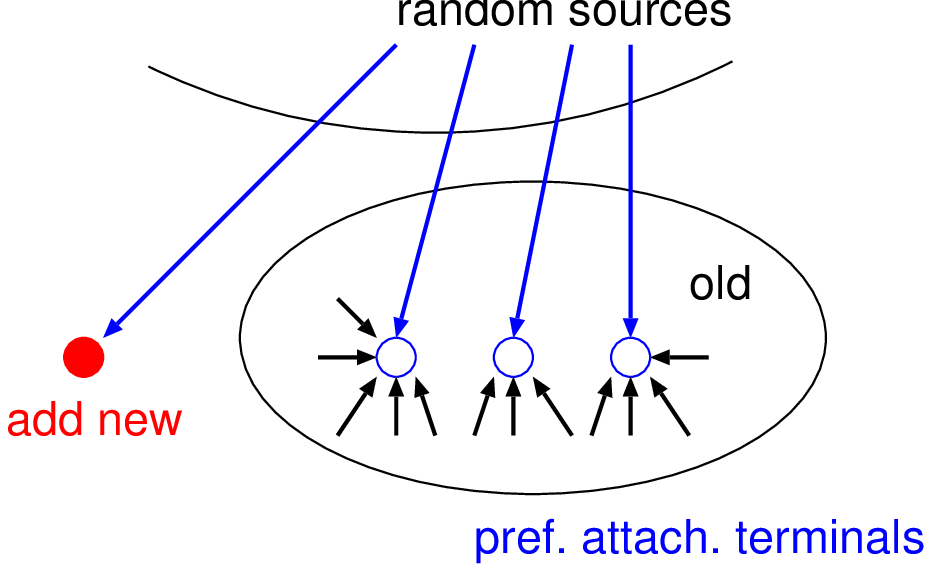} 
    \begin{center} (b) \end{center}
  \end{minipage} 
  \caption{Growing network models. (a) In the duplication-divergence
 model, duplication links 
 (red solid lines) between a new node (red circle) and all the neighbors
 (green circle) of a randomly chosen node (blue circle), and
 self-interaction (dashed line) are
 generated, but some links (magenta dashed line) are removed.
 (b) In the directed growing model, new links from randomly chosen sources
 to preferentially attached terminals emerge.}
  \label{fig_models}
\end{figure}

Let us consider the growing network models 
shown in Figs. \ref{fig_models} (a) and (b).
The following procedures 
are repeated until the network reaches to the required size $N$.

Dup: duplication-divergence model \cite{Vazquez03b}
\begin{enumerate}
  \item At each time step, 
	a new node $i'$ is added to the network.
  \item Simultaneously, a node $i$
	is randomly chosen, and new (undirected) links 
	between all the neighbors $j$ of $i$ 
	and the new node $i'$ are duplicated.
  \item With probability $q_{v}$,
	a link between $i$ and $i'$ is established
	(self-interaction).
  \item In the divergence process, 
	each duplicated link is removed 
	with probability $1 - q_{e}$.
\end{enumerate}
These local rules are biologically plausible \cite{Sole03}.
Note that larger $q_{v}$ enhances the assortativity of 
network generated by the above rules 
because the self-interaction means connecting a pair of 
nodes with similar degrees.
In other words, 
$q_{v}$ is a control parameter of the correlation.

Dir: directed growing model \cite{Dorogovtsev03}
\begin{enumerate}
  \item At each time step, 
	a new node is added and connected from 
	a randomly chosen node.
  \item Simultaneously, $m'$ new links emerge from randomly chosen nodes
	in the network.
  \item The terminals of the new links become attached to nodes
	chosen with shifted linear preference \cite{Krapivsky01}:
	a node with in-degree $k$ is chosen as the terminal of a new
	link with probability proportional to $k + w$, 
	where $w$ is a positive constant.
\end{enumerate}
The generation of new links 
includes the wiring between old nodes at each time step.
In addition, 
to keep the connectivity of network, the first procedure is modified 
from the probabilistic addition of new nodes  
in Ref. \cite{Dorogovtsev03}.
If a multi-link between already connected nodes or self-loop
is created, it is skipped 
in the directed growing model, while there is no such link 
in the duplication-divergence model.
In a sense of reality, 
the sender and the receiver of transmitted information or objects 
are distinguished from each other on a directed link.

\begin{table}[htb]
\begin{center}
\begin{tabular}{cc||cccc|ccc} \hline
Model & & $q_{v}$ & $q_{e}$ & $\gamma$ & $m'$ & $<k>$ & $K_{min}$ & $K_{max}$\\ \hline
Dup & Ass & 0.9 & 0.42 & - & - & 7.463 & 45 & 122\\
    & Unc & 0.3 & 0.48 & - & - & 7.356 & 55 & 159\\
    & Dis & 0.1 & 0.5  & - & - & 7.395 & 62 & 237\\ \hline
Dir & Ass & - & - & 3.0 & 7 & 7.33  &  80 & 142\\
    & Unc & - & - & 2.1 & 9 & 7.354 & 251 & 339\\ \hline
\end{tabular}
\vspace{2mm}
\caption{A set of parameters for the duplication-divergence and 
the directed growing models (denoted by Dup and Dir).
Unnecessary parameters are marked by hyphens.
The average degree of all nodes, 
the minimum and maximum degrees (in-degrees for directed links) 
of a hub node are measured over the 100 realizations of $N = 10^{3}$.
}
\label{table_param}
\end{center}
\end{table}

In the directed growing model, 
the rate equations for the in-degree distribution 
are written as 
\begin{equation}
  \frac{d N_{k}}{d \tau} = \frac{m'}{m + w} 
  \left[ (k-1+w) N_{k-1} -(k+w) N_{k} \right] + \delta_{k,1},
  \label{eq_deg_dir}
\end{equation}
where $m \stackrel{\rm def}{=} m' + 1$, 
$\gamma \stackrel{\rm def}{=} 1 + (m+ w)/m'$, 
$\delta_{k,1}$ is Kronecker's delta, 
the number of nodes with degree $k$ is denoted by 
$N_{k}(\tau) \sim n_{k} \times \tau$ 
as similar to Ref. \cite{Krapivsky01}, 
and we obtain the solution 
\begin{equation}
  n_{k} = \frac{(k-1+w) n_{k-1}}{k -1 + w + 1 +(m + w)/m'} 
  \sim k^{- \gamma}. \label{eq_deg_solution}
\end{equation}

\begin{figure}
  \begin{minipage}[htb]{.47\textwidth}
    \includegraphics[height=40mm, angle=0]{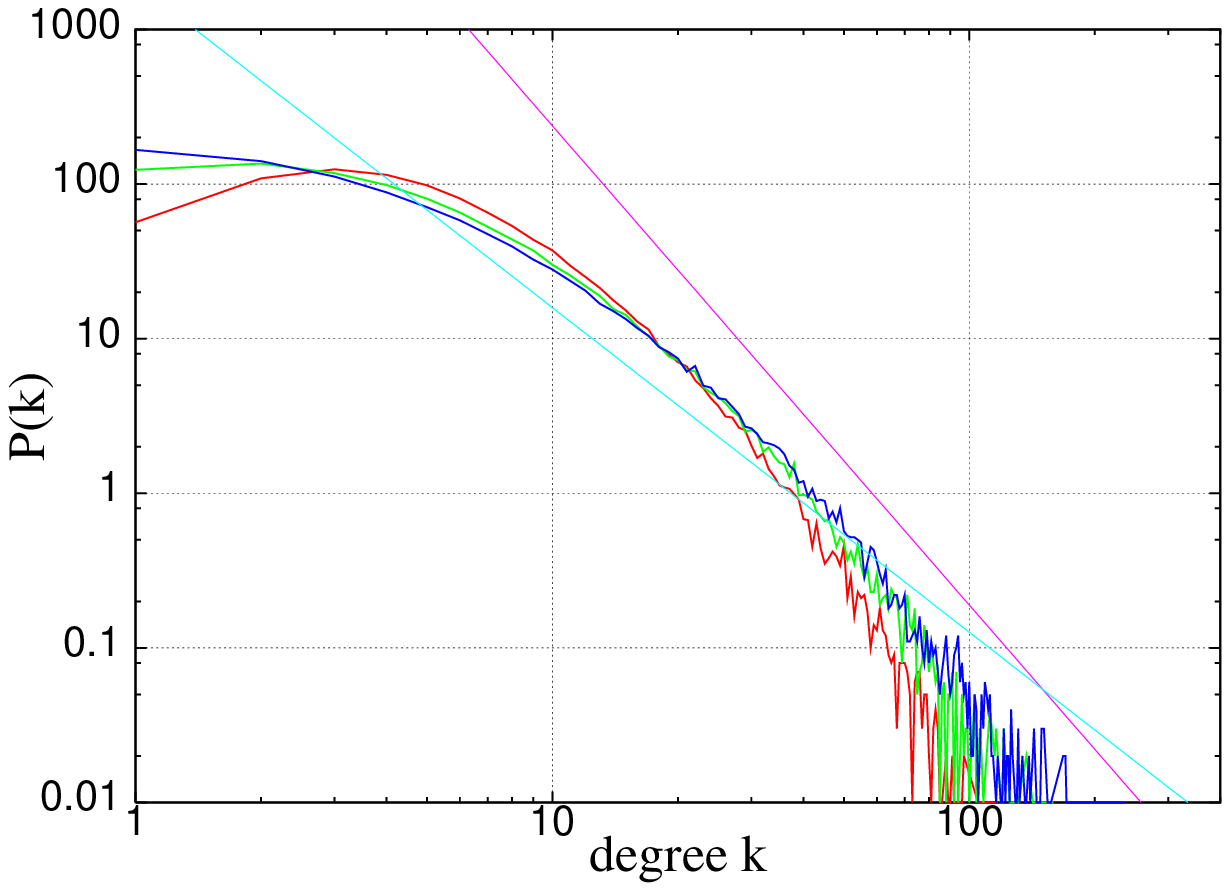} 
    \begin{center} (a) \end{center}
  \end{minipage} 
  \hfill 
  \begin{minipage}[htb]{.47\textwidth}
    \includegraphics[height=40mm, angle=0]{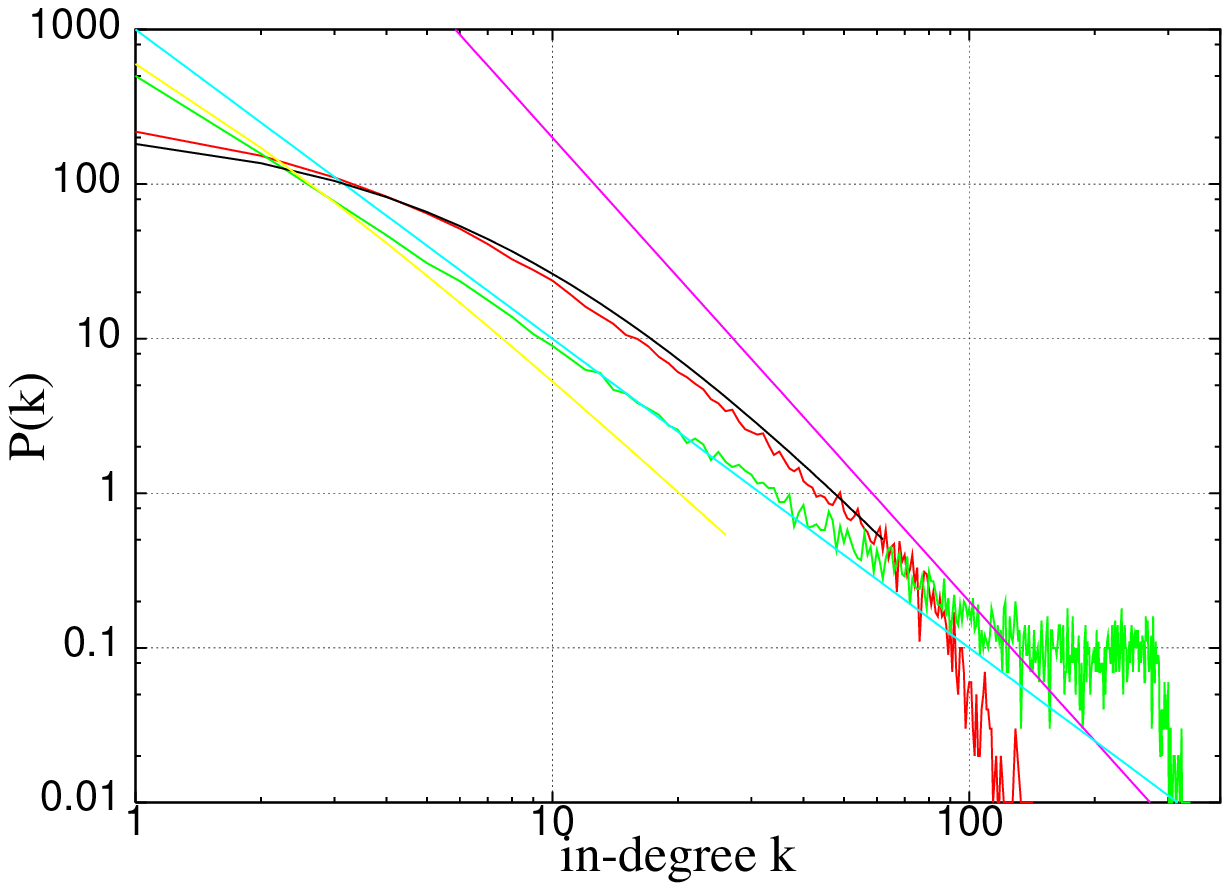} 
    \begin{center} (b) \end{center}
  \end{minipage} 
  \begin{minipage}[htb]{.47\textwidth} 
    \includegraphics[height=40mm, angle=0]{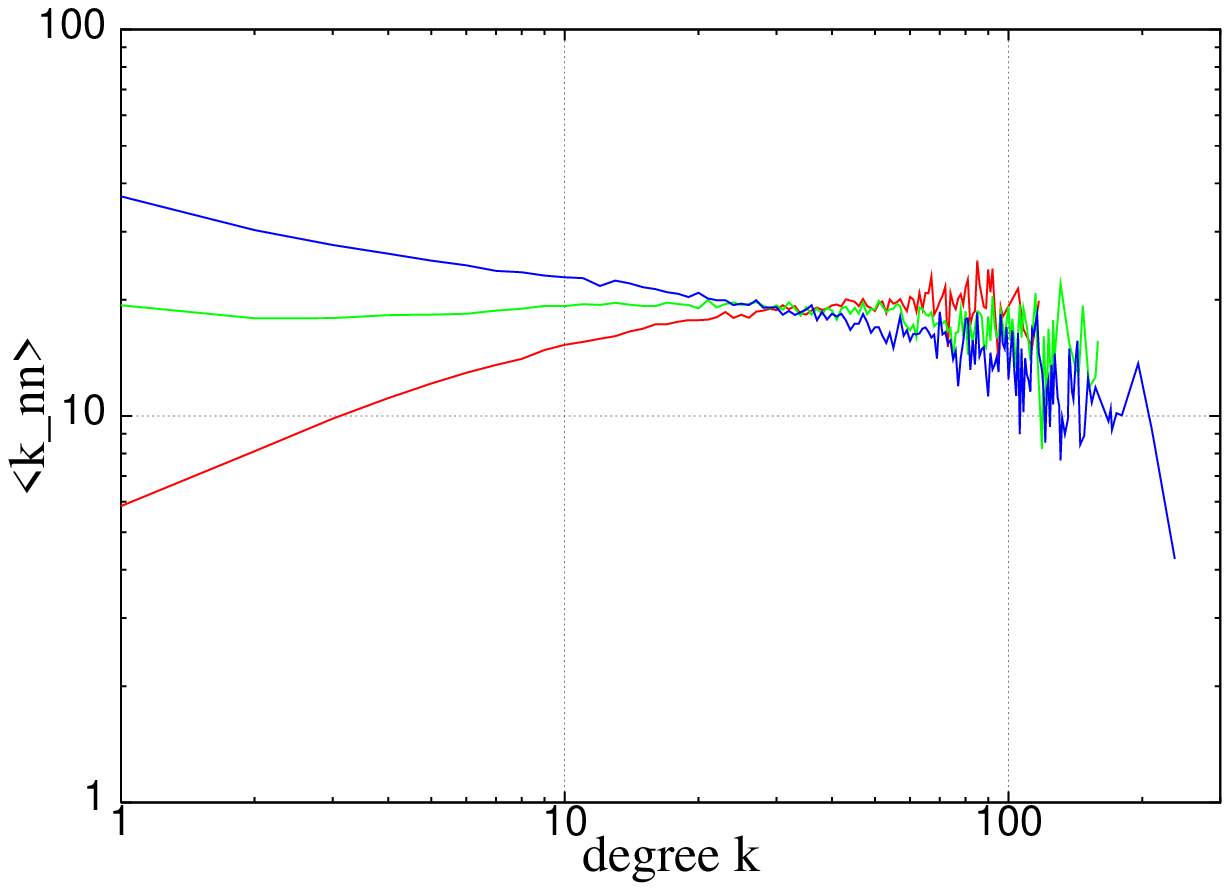} 
    \begin{center} (c) \end{center}
  \end{minipage} 
  \hfill 
  \begin{minipage}[htb]{.47\textwidth} 
    \includegraphics[height=40mm, angle=0]{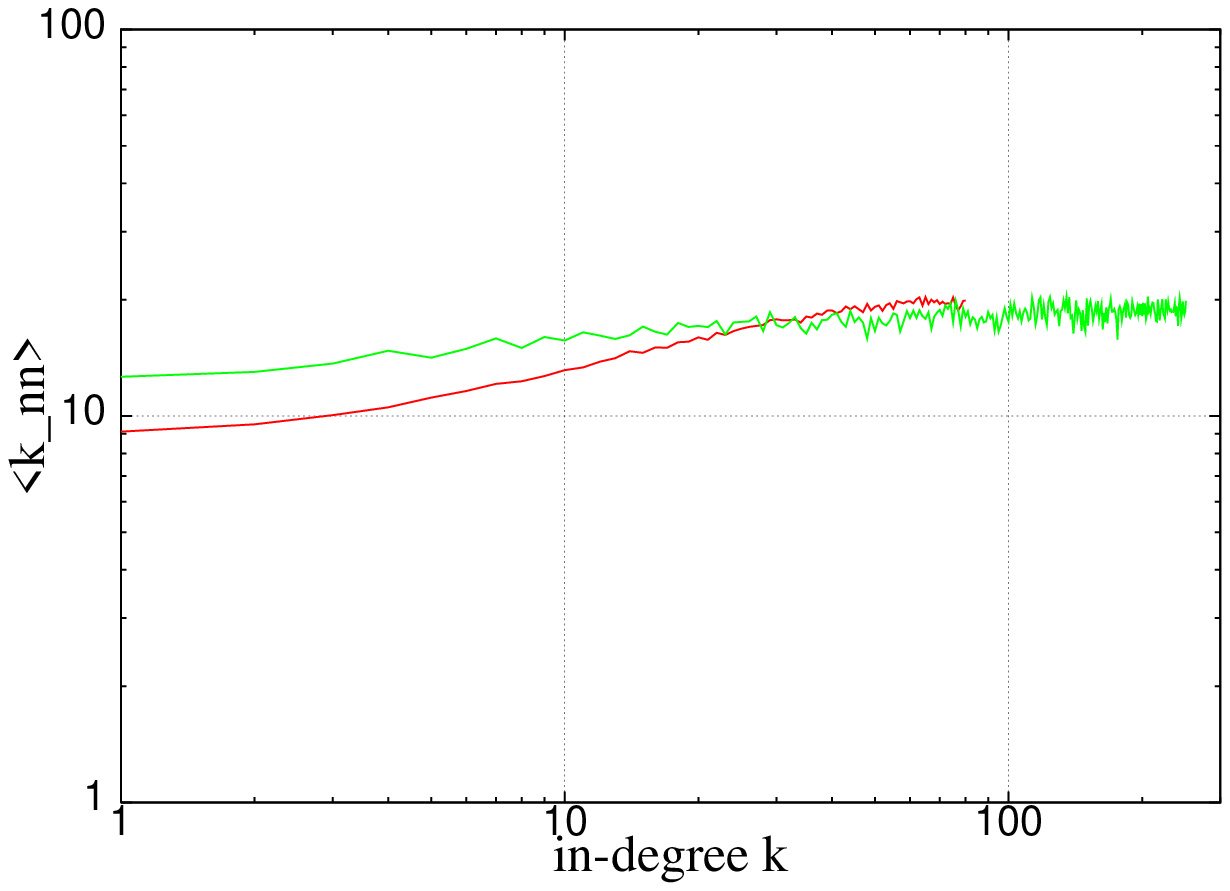} 
    \begin{center} (d) \end{center}
  \end{minipage}
\caption{Distributions of degree  and connectivity correlation. 
(a) $P(k)$ for the degree in Dup, (b) $P(k)$ for the in-degree in Dir, 
(c) $< k_{nn} >$ for the degree in Dup, 
(d) $< k_{nn} >$ for the in-degree in Dir.
The guide lines show power-law behavior of exponents 2.1 (cyan) and 3.0
 (magenta). In (b), the yellow and black lines 
for the analytical solutions of Eq
 (\ref{eq_deg_solution}) are well fitting the cases of $\gamma = 2.1$
 (green) and $3.0$ (red). In (c) and (d), 
the red, green, and blue lines 
clearly show the correlations corresponding to 
the cases of Ass, Unc, and Dis in Table \ref{table_param}.
The observations for very large degrees are statistically dropping and 
fluctuating. 
These are the averages over the 100 realizations.} 
\label{fig_deg_knn}
\end{figure}

The second column of Table \ref{table_param} shows 
the values of the parameters used in our simulation.
Ass, Unc, and Dis denote the assortative, uncorrelated,
and disassortative networks, respectively.
We regulate these values so that 
they produce similar average degrees $<k>$,
because it is obvious 
that the epidemic spreading becomes larger as the degrees increase.
Fig. \ref{fig_deg_knn} shows distributions of the degree 
$P(k) \stackrel{\rm def}{=} N_{k}/N$ and the connectivity correlation
$< k_{nn} > \stackrel{\rm def}{=} \sum_{l} l P(l|k)$,
where $P(l|k)$ is the conditional probability of the connections between
nodes with degrees $l$ and $k$ in Dup, 
or the connections from nodes with in-degree $l$ to those with $k$ in Dir. 
The degree distributions exhibit the power-law behavior:
the exponents vary as corresponding to the parameters $\gamma$ of 
Ass and Unc in Dir (Fig. \ref{fig_deg_knn} (b)),
while they are close to each other in Dup (Fig. \ref{fig_deg_knn} (a)).
The correlations are controlled between 
Ass and Dis in Dup (Fig. \ref{fig_deg_knn} (c)), while they are
controlled between 
Ass and Unc that is nearly uncorrelated with very weak correlations 
in Dir (Fig. \ref{fig_deg_knn} (d)).
If the connections are replaced by reciprocal ones after the
configuration of Dir, the degree distribution becomes exponential, and
the correlation disappears.

Next, we consider the SIR model 
in which individual nodes have three states: susceptible, infected, and
recovered/removed.
The densities of the nodes with degree (or in-degree in Dir) $k$
in the respective states 
are denoted by 
$s_{k}(t) = S_{k}(t)/N_{k}(t)$, $\rho_{k}(t) = I_{k}(t)/N_{k}(t)$, 
$r_{k}(t) = R_{k}(t)/N_{k}(t)$.
By definition, the normalization condition
$s_{k}(t) + \rho_{k}(t) + r_{k}(t) = 1$ holds at each time $t$.
Note that we distinguish the time-scales $\tau$ and $t$ for 
the evolution of network and the spreading of viruses.
After the construction of a network for the Dir or Dup model,
from an initial infected node, 
the epidemics is propagated by contacts between infected and susceptible
individuals (from infected nodes to susceptible nodes 
through directed links in Dir) 
at the rate $b$. 
The infected node is removed at the rate $\delta$. 
Once an individual gets infected and then recovers or removed, 
the state is never changed any more.
The microscopic stochastic simulation needs very expensive computation
for studying the epidemic properties, 
therefore a macroscopic mean-field approximation is useful.
We consider the spreading on the averages of randomly generated 
networks for each of Dup and Dir.

Following Ref. \cite{Boguna03}, 
the mean-field-like 
rate equations for the evolution of densities can be expressed as 
\begin{equation}
  \frac{d s_{k}(t)}{dt} = - b k s_{k}(t) \Theta_{k}(t),
   \label{eq_evol_s}
\end{equation}
\begin{equation}
  \frac{d \rho_{k}(t)}{dt} = - \delta \rho_{k}(t)
   + b k s_{k}(t) \Theta_{k}(t), 
 \label{eq_evol_rho}
\end{equation}
\[
 \frac{d r_{k}(t)}{d t} = \delta \rho_{k}(t), 
\]
where 
$\Theta_{k}$ in the r.h.s of Eqs. (\ref{eq_evol_s}) 
and (\ref{eq_evol_rho})
denotes the expectation of infection at degree $k$, 
\begin{equation}
 \Theta_{k}(t) \stackrel{\rm def}{=} \left\{
 \begin{array}{ll}
   \sum_{l = 1}^{k_{c}} \frac{l-1}{l} P(l|k) \rho_{l}(t) 
    & {\rm for \; Dup},\\
   \sum_{l = 1}^{k_{c}} P(l|k) \rho_{l}(t) 
    & {\rm for \; Dir},
  \end{array} \right. \label{eq_theta}
\end{equation}
the factor $(l - 1)/l$ is added taking 
account of the fact that one of the links is not available for
transmitting the infection because it was already used \cite{Boguna03}.
As applied to Fig. \ref{fig_deg_knn} (c) or (d),
the $P(l|k)$ 
is estimated from the average over 100 realizations for each parameter
specification presented in Table \ref{table_param}.
Since some degrees 
are missing in the range $K_{min} < k < K_{max}$ 
over the realizations,
we apply the cut-off value $k_{c}$ defined by $K_{min}$.

\begin{figure}
  \begin{minipage}[htb]{.47\textwidth}
    \includegraphics[height=40mm, angle=0]{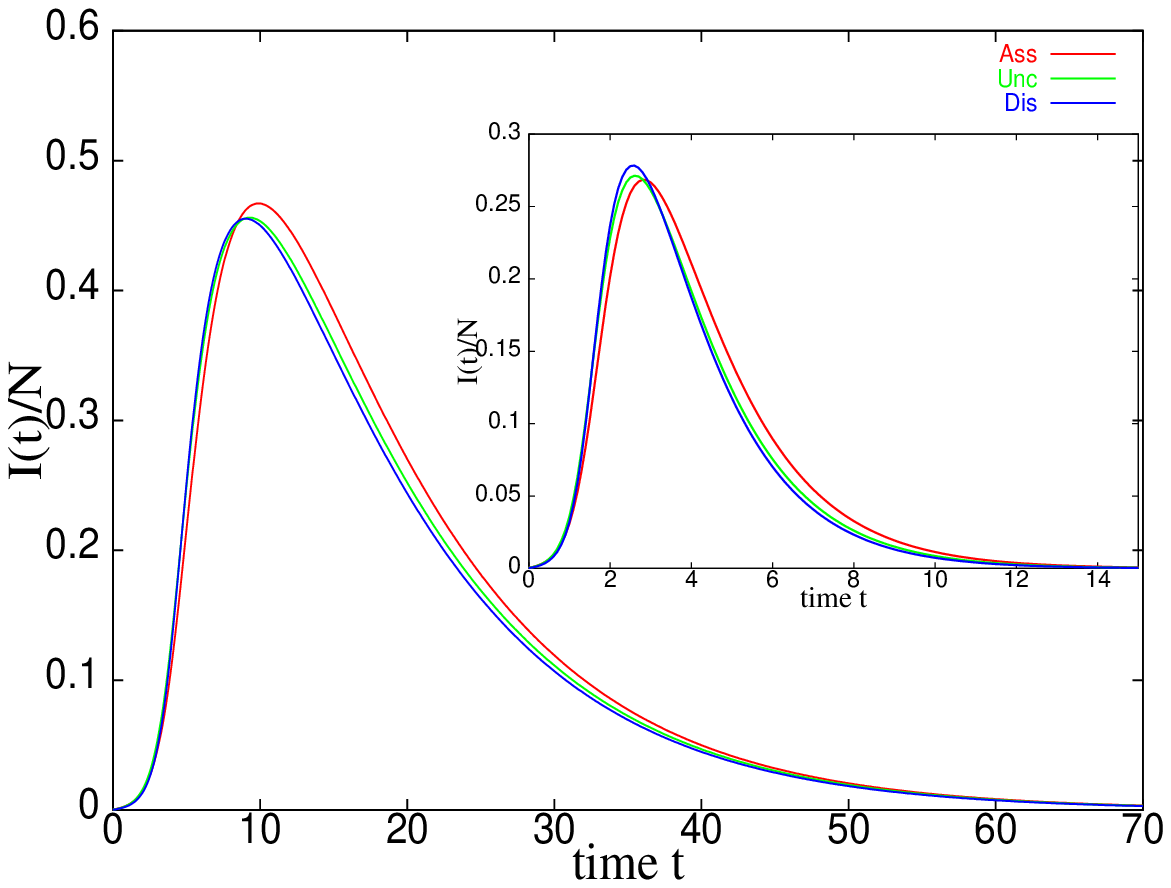} 
    \begin{center} (a) \end{center}
  \end{minipage} 
  \hfill 
  \begin{minipage}[htb]{.47\textwidth}
    \includegraphics[height=40mm, angle=0]{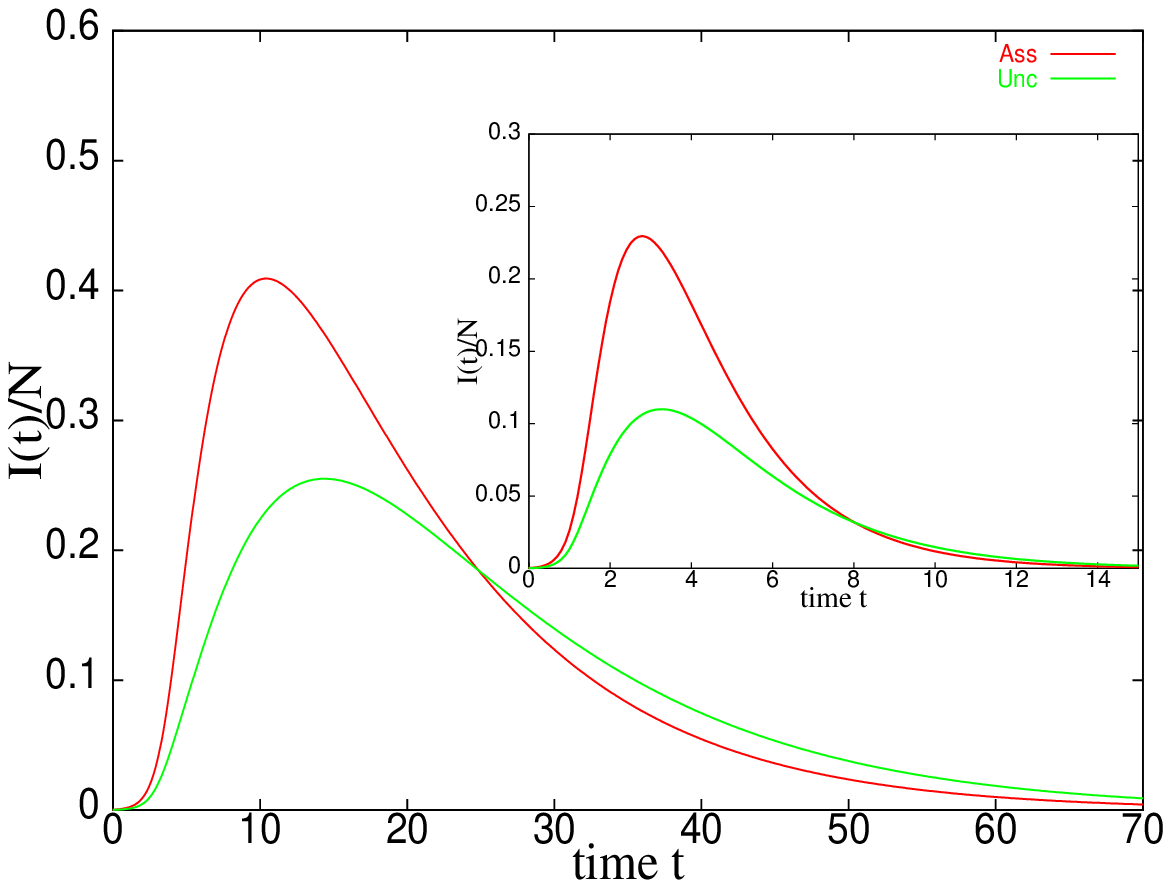} 
    \begin{center} (b) \end{center}
  \end{minipage} 
  \begin{minipage}[htb]{.47\textwidth} 
    \includegraphics[height=40mm, angle=0]{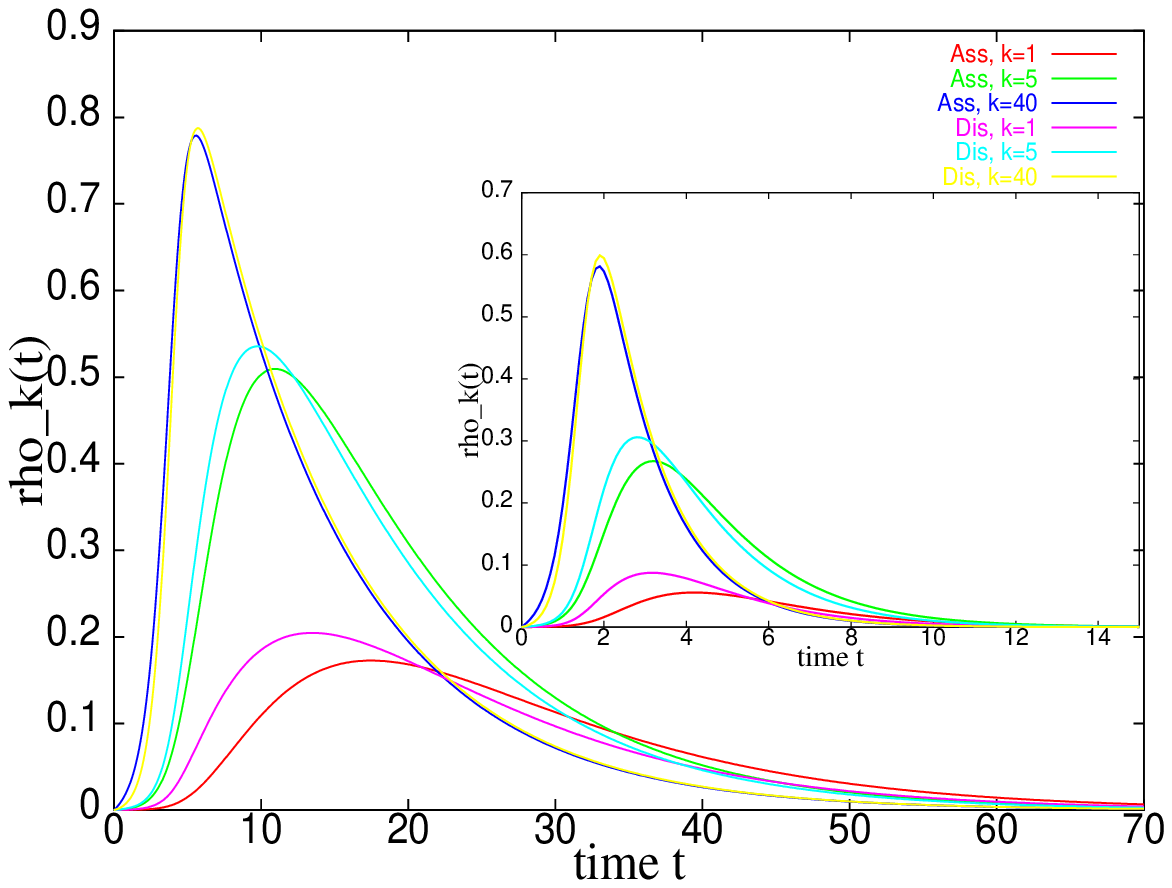} 
    \begin{center} (c) \end{center}
  \end{minipage} 
  \hfill 
  \begin{minipage}[htb]{.47\textwidth} 
    \includegraphics[height=40mm, angle=0]{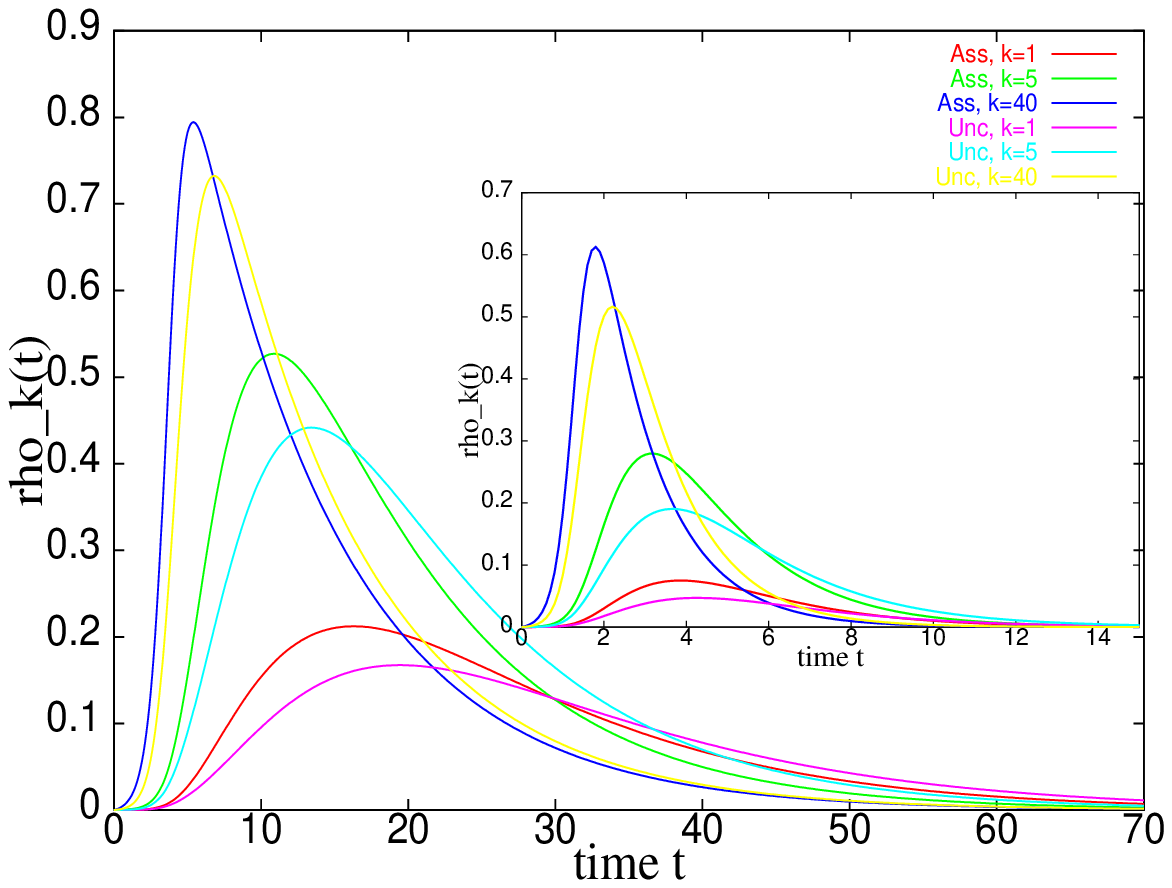} 
    \begin{center} (d) \end{center}
  \end{minipage}
\caption{Time evolutions of 
(a) the fraction of infected nodes $I(t) / N$ in Dup, 
(b) $I(t) / N$ in Dir, 
(c) each $\rho_{k}(t)$ for 
$k = 1, 5, 40$ in Dup, and (d) $\rho_{k}(t)$ in Dir
at $b = 0.1$ and $\delta = 0.1$.
Inset figures show similar behavior at $b = 0.3$ and $\delta = 0.5$
as some other example.
Note that the value of $I_{k}(t) = N_{k} \rho_{k}(t)$
is in the reverse order to $k$ 
because of $N_{k} \propto k^{- \gamma}$.} 
\label{fig_time_evol}
\end{figure}

We numerically investigate the dynamic behavior 
depends on the connectivity correlations 
by using the 4-th order Runge-Kutta method 
for Eqs. (\ref{eq_evol_s}) and (\ref{eq_evol_rho})
with a step width $\Delta t = 10^{-3}$.
In our simulation, we assume that an initial infection source is only 
on a hub
with the degree $k_{c}$.
Hence, for the other degrees $k \neq k_{c}$ 
we set $s_{k}(0) = 1$ and $\rho_{k}(0) = r_{k}(0) = 0$.
This assumption is natural since the hub is much more vulnerable against
infection through the active communications with the outside.

\begin{figure}[htb]
  \begin{minipage}[htb]{.47\textwidth}
    \includegraphics[height=55mm, angle=0]{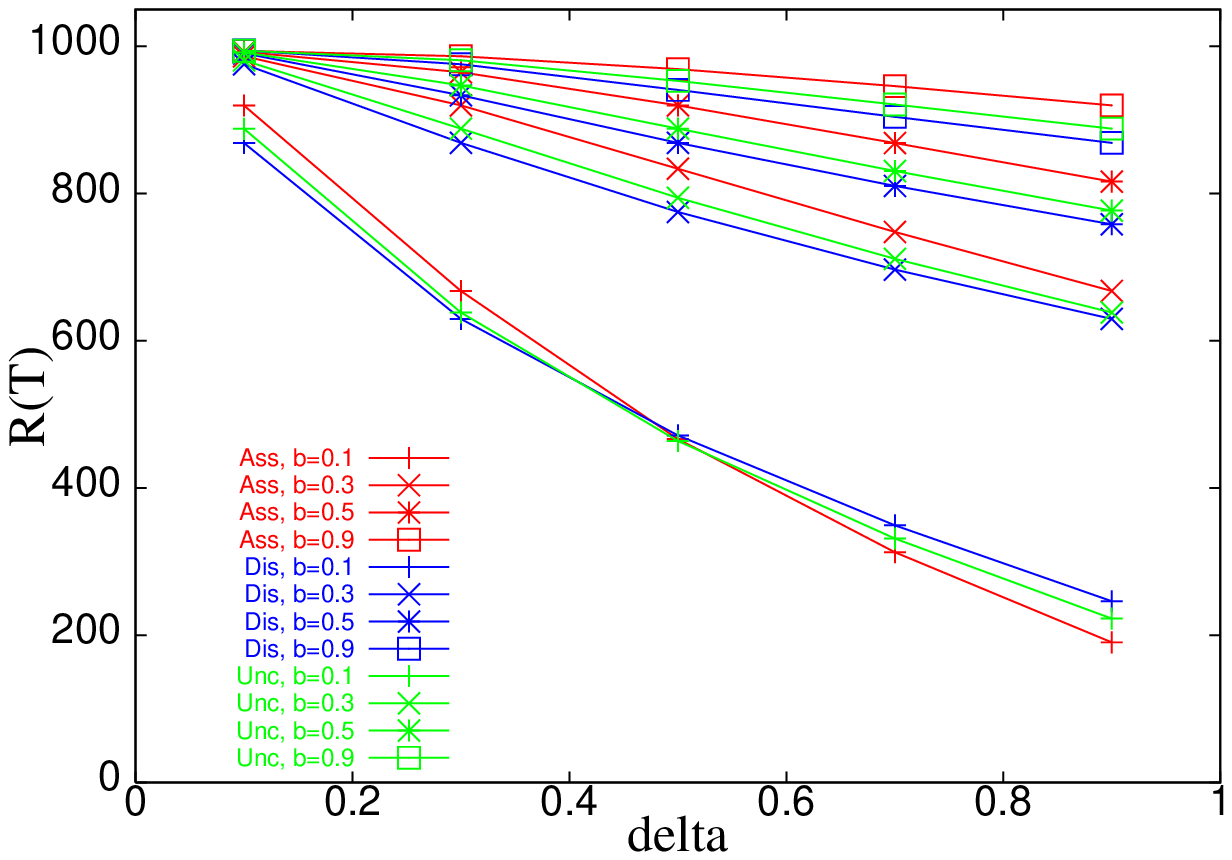}
    \begin{center} (a) \end{center}
  \end{minipage} 
  \hfill 
  \begin{minipage}[htb]{.47\textwidth}
    \includegraphics[height=55mm, angle=0]{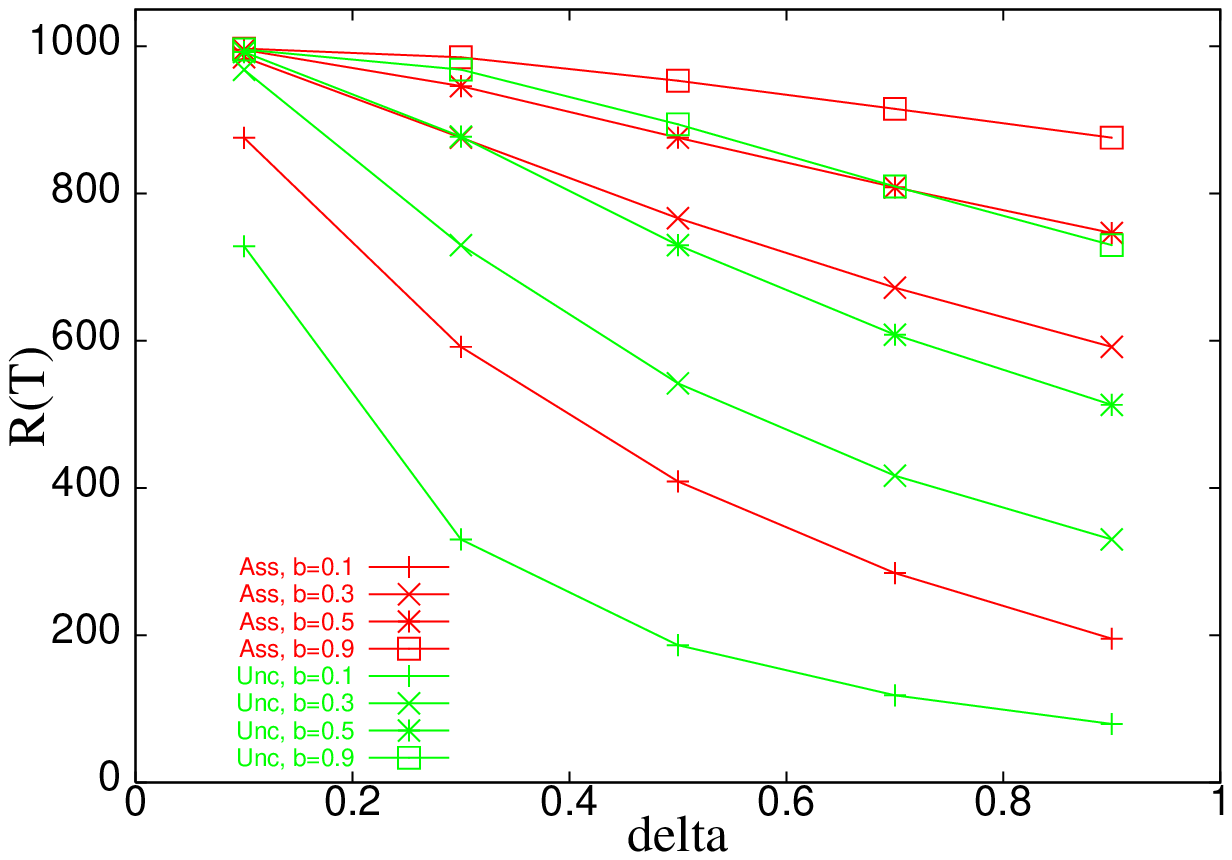} 
    \begin{center} (b) \end{center}
  \end{minipage} 
\caption{The epidemic incidence $R(T)$ as a function of the immune rate
 $\delta$ in (a) Dup, and (b) Dir. The red, green, and blue lines
 are corresponding to the cases of Ass, Unc, and Dis 
 at $b = 0.1, 0.3, 0.5, 0.7$ marked by 
 the plus, cross, asterisk, and open square, respectively.} 
\label{fig_total_R}
\end{figure}

Figs. \ref{fig_time_evol} (a) and (b) show
the fraction of the infected nodes 
$I(t)/N = \sum_{k} I_{k} / N$, and we see 
in the case of Ass (red line) it is slightly more insistent 
(Fig. \ref{fig_time_evol} (a) for Dup)
or larger (Fig. \ref{fig_time_evol} (b) for Dir)
than that in the case of Dis (blue line) and Unc (green line).
The densities of the infected nodes $\rho_{k}(t)$ in 
Figs. \ref{fig_time_evol} (c) and (d)
show a property similar to that of $I(t)/ N$.
The value of $\rho_{k}(t)$ becomes larger as the degree $k$ increases,
therefore the infection from the nodes with high degrees 
are dominant in the $\Theta_{k}$ of Eq. (\ref{eq_theta})
in particular through assortative connections.
Fig. \ref{fig_total_R} shows the epidemic incidence 
$R(T) \stackrel{\rm def}{=} \sum_{k} R_{k}(T)$,
which denotes the number of transitions from the total infected nodes.
We set $T = 100$ taking the convergent time 
in Fig. \ref{fig_time_evol} into account.
The incidence is higher in the assortative networks (red line)
than that in uncorrelated (green line) or disassortative networks (blue
line), except at $b = 0.1$ and $\delta = 0.7, 0.9$ in Dup.
The exceptions suggest that 
the immune rate $\delta$ is an important factor for small $b$.
In other words, if we set $\delta' = 1$ by a variable change of time
$t$ for Eqs. (\ref{eq_evol_s}) and (\ref{eq_evol_rho}), 
we must use $b' = b / \delta$, which may be larger than 1, 
instead of $0 < b < 1$.
These differences according to the types of correlations 
remarkably appear in Dir with smaller $b$.
Furthermore, 
similar results are also obtained in the directed growing model 
based on preferentially attached terminals and sources by 
the in- and out-degrees, respectively.

In summary, we have numerically 
found different spreading properties 
on SF networks according to the connectivity correlations 
estimated from the averages of the growing models
\cite{Dorogovtsev03}\cite{Vazquez03b}.
The differences are remarkable in the directed model.
The results suggest that assortative connections enhance the 
epidemic spreading, 
and also they could contribute to improve the efficiency of 
information delivery. In contrast, the disconnections or setting 
gatekeepers between nodes with similar degrees becomes a local defense
strategy such as acquaintance immunization \cite{Cohen03}
for the spreading of viruses.
Further works will be carefully considered 
for comprehending the effects of correlations on the spreading.

\end{document}